\documentclass[journal=jpcafh,manuscript=article]{achemso}
\usepackage[version=3]{mhchem} 
\usepackage[T1]{fontenc}   
 
\author{Maciej Bartosz Kosicki}
\email{mackos@fizyka.umk.pl}
\affiliation[NCU Institute of Physics]
{Institute of Physics, Faculty of Physics, Astronomy and Informatics, Nicolaus Copernicus University, Grudzi\k{a}dzka 5, 87-100, Toru\'n, Poland}
\author{Dariusz K\k{e}dziera}
\email{teodar@chem.uni.torun.pl} 
\affiliation[NCU Department of Chemistry]
{Department of Chemistry, Nicolaus Copernicus University, 7 Gagarin Street, 87-100 Toru\'n, Poland}
\author{Piotr Szymon  \.{Z}uchowski}
\email{pzuch@fizyka.umk.pl}
\affiliation[NCU Institute of Physics]
{Institute of Physics, Faculty of Physics, Astronomy and Informatics, Nicolaus Copernicus University, Grudzi\k{a}dzka 5, 87-100, Toru\'n, Poland}

\title[ Chemical reactions of cold SrF and CaF molecules with Alkali-Metal and Alkaline-Earth-Metal atoms]
  {\textit{ Ab Initio} Study of Chemical Reactions of Cold SrF and CaF Molecules with  Alkali-Metal and Alkaline-Earth-Metal Atoms: the Implications for Sympathetic Cooling}

\begin{document}
 \begin{abstract} 
We investigate the energetics of the atom exchange reaction in the SrF+alkali-metal atom and CaF+alkali-metal atom systems. Such reactions are possible only for collisions of SrF and CaF with the lithium atoms, while they are energetically forbidden for other alkali-metal atoms.  Specifically, we focus on SrF interacting with Li, Rb, and Sr atoms and use  {\it ab initio} methods to demonstrate that the SrF+Li and SrF+Sr reactions are barrierless. We present potential energy surfaces for the interaction of the SrF molecule with the Li, Rb, and Sr atoms in
their energetically lowest-lying electronic spin states. The obtained potential energy surfaces are deep and exhibit profound interaction anisotropies.  
We predict that the collisions of SrF molecules in the rotational or Zeeman excited states most likely have a strong inelastic character. 
We discuss the prospects for the sympathetic cooling of SrF and CaF molecules using ultracold alkali-metal atoms.

\end{abstract}

\section{Introduction}
Polar molecules in the ultracold regime have been extensively studied in recent years. They became a unique platform for precise tests of fundamental physics~\cite{Kozlov:2002, Hudson:2011, Tarbutt:2013}, quantum simulations ~\cite{Bloch:2012,DeMille:2002,BuechlerPRL07} and fundamental studies of chemical reactions close to the thresholds, which to date was realized for the KRb molecule~\cite{Ospelkaus:2009, deMiranda:2011}.
Heteronuclear molecules have been already formed in their rovibrational X${}^1 \Sigma^+$ ground states for the following species: KRb \cite{Ni:KRb:2008}, RbCs~\cite{Cornish:2014,Takekoshi:2014}, NaK \cite{Zwierlein:2015}, and NaRb \cite{Wang:2016}.
These molecules were obtained from ultracold atoms by first preparing them in weakly bound Feshbach states and transferring them to the electronic ground state via stimulated Raman adiabatic passage (STIRAP)\cite{Vitanov:2017}.

Apart from alkali-metal dimers,  there is a great interest in cooling and trapping of the paramagnetic molecules in the X${}^2 \Sigma^+$ electronic ground state. It was found by Di Rosa\cite{DiRosa2004} that some of these molecules  have  closed cycling transitions due to strongly diagonal 
Franck-Condon factors and thus, can be laser cooled. A number of experiment is focused now on the laser cooling of such molecules as, for example, SrF, CaF, YbF, and YO. \cite{Barry:2014,Zhelyazkova2014,Lim:2015, Tarbutt:2013, Chae:2017}.  More recently, Norrgard \textit{et al.} \cite{Norrgard:2016} demonstrated the 3D magneto-optical  trapping 
of SrF molecules below milikelvin. However, to produce the samples of such molecules with the temperatures on the order of $\mu$K is still a formidable task.

One promising way to make molecules even colder is sympathetic cooling, in which temperature of one species is reduced by thermal contact with another, much colder collision partner.
  The possibilities for sympathetic cooling of polar molecules with alkali metal atoms or alkaline-earth metal atoms were recently explored for various systems, such as, NH~\cite{Soldan2004,Campbell:2007,Soldan:2009,Wallis2011,Hummon:2011}, LiH~\cite{Skomorowski:2011}, and CaH  \cite{Tscherbul:2011}. Very recently, collisions of CaF with the Li and Rb atoms were investigated by Frye and co-workers \cite{Frye:2016} as well as Lim \textit{et al.}~\cite{Lim:2015}. In the former the Feshbach resonances in the low-collision regime were studied using {\it ab initio} interaction potential for spin-stretched state, while in  the latter the Lennard-Jones potential was used to explore the possibility of sympathetic cooling of CaF in the absolute ground state.
  
Up to date, very little is known about the interactions of monofluoride molecules, such as SrF,  which are highly relevant for ongoing experiments. Chemical pathways in reactions of SrF molecules were investigated theoretically by Meyer and  Bohn~\cite{Meyer:11}. These authors found that the SrF+SrF reaction is barrierless and leads to the SrF$_2$ molecule and Sr atom. More recently Qu\'{e}m\'{e}ner and Bohn showed that X${}^2 \Sigma^+$ molecules can be shielded from inelastic and reactive collisions in the first rotational state~\cite{Quemener:2016}.  

The main motivation for the current work is to explore the interactions of the SrF and CaF molecules with atom that can be laser cooled: alkali-metal atoms and alkaline-earth metal atoms.   To this end, we first investigate possible chemical reactions that might happen in the ultracold mixtures of the SrF and CaF molecules and alkali-metal atoms to figure out which  could be used to sympathetically cool the absolute ground state molecules, without endangerment of chemical reaction at low collision energy. We employ high-level quantum chemistry methods  to study in detail SrF molecule and characterize its spectroscopic and electric properties. Secondly, we focus in details on the interactions in SrF+Li, SrF+Rb and SrF+Sr systems in the entrance and exit channels of the chemical reactions in Jacobi coordinates. We characterize  the interaction anisotropies, and determine the isotropic and anisotropic van der Waals coefficients for the long-range part of the interaction potential. Finally, we study the  potential energy surface for the atom exchange reaction in the internal coordinates.  The last section concludes our work and outlines further directions of research.

\section{Methodology}
Spectroscopic constants for diatomic species in the electronic ground state were obtained using the coupled cluster method with single, double and noniterative triple excitations [CCSD(T)] \cite{Hampel:92, Knowles:94}. 
For the Li, Na and F atoms the augmented core-valence,  correlation-consistent basis sets were used (aug-cc-pCV5Z, aug-cc-pCVQZ), while to for the K, Rb, Cs, Ca, and Sr atoms we employed the small-core relativistic energy-consistent pseudopotentials of Stuttgart group (ECP28MDF) with tailored valence basis set truncated to \textit{spdfgh} and \textit{spdfg} orbitals in the uncontracted form \cite{Lim:2005, Lim:2006, Prascher:2011, Woon:95}.  We performed two-point extrapolation to complete basis set limit (CBS) \cite{Helgaker:2000} and  the counter-poise correction scheme was applied to compensate the basis set superposition error (BSSE)\cite{Boys:70} in the supermolecular interaction energy.
Moreover, to improve description of the dispersion interactions we added \textit{spdf}  midbond functions located in the center of mass between two monomers \cite{TaoPan:91}. 

Electronic structure calculations for the SrFLi, SrFRb, and \ce{Sr2F} trimers were performed in few steps. 
First, the Intrinsic Reaction Coordinate method at second order of M{\o}ller-Plesset level of theory (IRC-MP2)\cite{Fukui:1970,Schmidt:1985, Head-Gordon:1988,Frisch:1990} was used to study the pathway between stationary points on the potential energy surface. For the IRC calculations we employed a double-$\zeta$ quality  basis sets.
Second, we reoptimized the IRC-MP2 geometries using the CCSD(T) approach with energy threshold $10^{-7}$ a.u. of energy.
In this case the basis sets was truncated to \textit{spdf} functions in the aug-cc-pCVQZ basis sets for Li and F atoms as well as ECP28MDF pseudopotentials were employed to describe heavier atoms. 
We also computed stationary points on the potential energy surface for remaining fluorine-strontium-alkali-metal-atom, fluorine-calcium-alkali-metal-atom, \ce{Ca2F} trimers in the ground states using CCSD method \cite{Eckert:1997}. For the calcium atoms we employed the ECP10MDF pseudopotential\cite{Lim:2006}.  

 In the next step, we calculated  potential surfaces corresponding to the ${}^1A'$ (for SrF+Rb and Li) and ${}^2A'$ (for SrF+Sr) electronic states in the C${}_s$ symmetry point group: they were obtained in the Jacobi coordinate system in which we used the angle between the center-of-mass (COM) of a molecule and atom, axis of a molecule, and the COM-atom distance.  
The reaction pathways were studied as a function of distances between the metal and F atoms: in such case the Sr-F-X angle was fixed, with X being either the Li, Sr, or Rb atom. 
 
We also examined the minimum energy path between the global and local minimum 
on potential energy surfaces using the CCSD(T) approach (all possible electrons correlated). However, the CC method cannot properly work for the closed-shell system dissociating into open-shell fragments, which is the case in the SrF+ alkali-metal-atom in the lowest spin state.
 Hence, the full {\it ab initio} calculations were performed for the intermolecular distances up to 6, 7, and 10 \AA{} for SrF+Li, SrF+Rb, and SrF+Sr, respectively. For these distances the CC method was possible to converge and the T1-diagnostics\cite{Lee:1989} (which monitors the reliability of CC calculations) was smaller than 0.02. 

All coupled-cluster and IRC calculations were performed using the MOLPRO package \cite{MOLPRO_brief:2015} and the Q-Chem programs \cite{QCHEM4}.

\section{Results and discussion}
\subsection{Spectroscopic properties of the SrF molecule} 
First, we investigate molecular properties of SrF to asses the performance of the CCSD(T) method.
Spectroscopic properties of the SrF molecule in the  X${}^2\Sigma^+$ electronic ground state configuration obtained form {\it ab initio} calculations and experiments are collected in Table \ref{tbl:srf}. 
One can see that spectroscopic constants such as the equilibrium distance, R${}_e$, and the potential energy depth, D${}_e$, calculated in this work are in excellent agreement with the available experimental data\cite{Engelke:1979, Huber}. 
The harmonic frequency, $\omega_{0}$ and the rotational constant, B${}_{0}$ computed by means of the discrete variables representation approach (DVR) with the basis set introduced in the Ref.~\cite{ColberMiller:1992} also agree well with the reference data \cite{Bernath:1996}. 

The ground state of the SrF molecule exhibits a very strong charge transfer: it is known that for such systems asymptotic ground state is Sr+F, which is below the Sr$^+$+F$^-$ state~\cite{Meyer:11}. 
Using a very simple model for the harpoon reaction (see, for example~\cite{Soldan2004}) it is possible to estimate the position of avoided crossing between neutral Sr+F and ion-pair states Sr$^+$+F$^-$ as   $\frac{e^2}{4\pi\epsilon_0 \Delta E_x}= 6.3$ \AA{}, where $\Delta E_x$ is the difference between the ionization potential of Sr and electron affinity of F atoms. With the supermolecular CCSD(T) approach one can obtain the fragments of ion-pair and neutral interaction energies far from the avoided crossing and  interpolate them to find their the anticrossing position at 6.56 \AA{}.

The electric properties in equilibrium geometry, dipole moment and polarizabilities  obtained using the finite-field approach also are reported in Table \ref{tbl:srf}.
The permanent electric dipole moment $\mu$ is in very good agreement with the experimental results of Ernst and coworkers \cite{Ernst:1985}. 
As can be seen in Table \ref{tbl:srf}, the average static polarizability $\bar{\alpha}$ is about a three times larger than its anisotropy  $\Delta{\alpha}$. 
This is much less than in case of typical alkali dimers\cite{Deiglmayr:2008pol,Zuchowski:2014}. It is also worth to comment that the parallel component of the polarizability (125 a.u., not shown in Table \ref{tbl:srf}) agrees very well with the value given by Meyer and Bohn \cite{Meyer:11} (126 a.u.).  

To conclude this part, the CCSD(T) method correctly reproduces spectroscopic properties of the SrF molecule and is a proper choice for further calculations of potential energy surfaces. 

\begin{table}
  \caption{Properties of the ${}^{88}$Sr${}^{19}$F molecule in the electronic ground state X${}^2\Sigma^{+}$. Spectroscopic constansts D${}_0$, D${}_e$, R${}_e$, $\omega_{0}$, B${}_{0}$  were obtained using the extrapolation of basis set limit in  the CCSD(T) method. Electric properties $\mu$, $\overline{\alpha}$, $\Delta\alpha$ were obtained using the finite-field approach for molecule with fixed equilibrium distance which are taken from literature \cite{Huber}. The reference data correspond to the experimental values\cite{Engelke:1979,Huber, Ernst:1985,Bernath:1996} .}
  \label{tbl:srf}
  \begin{tabular}{lll}
    \hline
    & This work & Reference   \\
    \hline
    D${}_0$ (cm${}^{-1}$) & 44797 & -   \\
    D${}_e$ (cm${}^{-1}$) & 45047 & 45290 \cite{Engelke:1979}  \\
    R${}_e$ (\AA) & 2.081 & 2.075 \cite{Huber}\\
    $\omega_{0}$ (cm${}^{-1}$)& 500.25 & 501.96493 \cite{Bernath:1996} \\
    B${}_{0}$ (cm${}^{-1}$)& 0.248 & 0.24975 \cite{Bernath:1996} \\
    $\mu$ (D) & 3.46 & 3.4676 \cite{Ernst:1985}\\
    $\overline{\alpha}$ (a.u.) & 170.05 & - \\
    $\Delta{\alpha}$ (a.u.) & 66.35 & - \\
    \hline
  \end{tabular}
\end{table}

\subsection{Energetics of chemical reactions of alkali-metal atoms with SrF and CaF molecules} 
{
As it was mentioned in the Introduction, SrF or CaF molecules might be produced in even lower temperature by sympathetic cooling through collisions  with ultracold, laser-cooled atoms. 
Such experiment might be realized by overlapping (optically or magnetically) trapped atoms and cold cloud of monofluoride molecules in MOT.
The key information which can be provided by {\em  ab initio} electronic structure theory, is whether the chemical reactions between atoms and such molecules might occur
or not. To this end we first focus on the atom exchange reactions: SrF+X$\to$ XF+Sr and CaF+X$\to$ XF+Ca, where X is the alkali-metal atom.
The energy differences for these reactions computed as the difference of zero point energies D${}_0$ of appropriate monofluoride molecules are collected in Table \ref{tbl:deltaE}. 

As can be seen in Table \ref{tbl:deltaE}, SrF+Li and CaF+Li reactions are only systems which are chemically reactive.
In the case of the SrF+Li reaction, the amount of energy (3551 cm${}^{-1}$) is sufficient to populate first five vibrational energy levels of the LiF molecule in the electronic ground state.
For the remaining alkali-metal atoms the reaction with SrF or CaF molecules is energetically forbidden. 
This may be explained by fact that the dissociation energy for the LiF molecule is significantly larger compared with other monofluorides. 

\begin{table}
  \caption{  The energy differences for the exchange reactions in  ${}^{88}$SrF and ${}^{40}$CaF collisions with alkali 
metal atoms obtained from the formula: $\Delta E = D_{0}^{XF} -D_{0}^{YF}$, where $X$ refers to the Sr or Ca atom and $Y$ is the alkali-metal atom. Dissociation energies of electronic ground states correspond to the CBS limit in the CCSD$($T$)$ method. For the CaF molecule the  $D_0$ is taken the from literature~\cite{Huber}.}
  \label{tbl:deltaE}
  \begin{tabular}{ll}
    \hline
    chemical reaction &  $\Delta$E (cm${}^{-1}$)  \\
    \hline
    ${}^{}$Sr${}^{}$F + ${}^7$Li & -3551  \\
    ${}^{}$Sr${}^{}$F + ${}^{23}$Na & 4796 \\
    ${}^{}$Sr${}^{}$F + ${}^{39}$K & 3753 \\
    ${}^{}$Sr${}^{}$F + ${}^{85}$Rb &3566 \\
    ${}^{}$Sr${}^{}$F + ${}^{133}$Cs & 2029 \\
    \hline
    ${}^{}$Ca${}^{}$F + ${}^7$Li & -4440  \\ 
    ${}^{}$Ca${}^{}$F + ${}^{23}$Na & 3907 \\
    ${}^{}$Ca${}^{}$F + ${}^{39}$K & 2864 \\
    ${}^{}$Ca${}^{}$F + ${}^{85}$Rb &2677 \\
    ${}^{}$Ca${}^{}$F + ${}^{133}$Cs & 1140 \\\hline
  \end{tabular}
\end{table}

\subsection{Equilibrium geometries and local minima in the SrFX and CaFX trimers}

It is interesting to inspect the stationary points on the potential energy surface of the electronic ground state for following systems: SrFX, CaFX, \ce{Sr2F} and \ce{Ca2F}, where X is the alkali-metal atom.
Equilibrium geometries optimized in the CCSD method are collected in Table ~\ref{tbl:optg}. 
It can be seen that the angle between monomers increases with the mass of the alkali-metal atom in trimer starting from Na. On the other hand the bond length between Sr and Ca with F atoms exhibits only a small change compared to isolated monofluoride molecule.
For the \ce{Sr2F} and \ce{Ca2F} trimers the optimal geometry corresponds to the  C$_{2v}$ symmetry and the bond lengths in these systems are slightly longer than in isolated molecules. 

In present study we are particularly focused on the interactions in SrFLi, SrFRb and \ce{Sr2F} trimers. For these systems we present global- and local minima, and saddle points optimized using the CCSD(T) method (see Table~\ref{tbl:lowspin_points}).
Results of the CCSD(T) calculations for the global minima agree well with those obtained from the CCSD approach.
In the case of the SrFLi system in their equilibrium geometry, the distance between the F and Li monomers  (1.70 \AA) is close to the equilibrium distance of the LiF molecule (1.56 \AA) which is significantly smaller than the bond length of the SrF in the SrFLi trimer (2.27 \AA). In contrast,  the intermolecular distance in the RbF dimer is about 15\% longer than that in the SrF dimer and significantly longer than in the equilibrium distance of the RbF molecule (2.60 compared to 2.27 \AA). 
The local minima  on the potential energy surface for SrFLi and SrFRb trimers are observed for the linear geometry. 
The optimal geometry for the \ce{Sr2F} system is a isosceles triangle (the C${}_{2v}$ symetry point group) with bond lengths  comparable to the equilibrium distance in diatomic SrF molecule. The geometry of the saddle point of \ce{Sr2F} differs from these of SrF-alkali-metal-atom trimers: there is a strongly non-symmetric configuration with bond lengths between F and Sr atoms equal 4.58 and 2.10 \AA{}, respectively.  

\begin{table}
    \caption{Geometries of local minima of the electronic ground state optimized in the CCSD approach for SrFX and CaFX trimers where X is the alkali-metal atom or Sr/Ca atom. The bond lengths are given in \AA{}. }
  \label{tbl:optg}
  \begin{tabular}{lllll}
    \hline
     System & R${}^{}_{SrF}$ & R${}^{}_{FX}$ & $\theta_{SrFX}^{}$   \\
    \hline  
    SrFLi   &2.26 &1.67  &106.16  \\
    SrFNa   &2.21 &2.07  &104.49    \\
    SrFK    &2.22 &2.36  &112.51 \\
    SrFRb   &2.22 &2.477  &115.09  \\
    SrFCs   &2.22 &2.587  &120.58  \\
    \hline
    CaFLi   &2.13 &1.67  &105.47  \\
    CaFNa   &2.09  &2.07  &103.61 \\ 
    CaFK    &2.10 &2.36  &111.63  \\
    CaFRb   &2.10 &2.47  &114.41  \\
    CaFCs   &2.10 &2.59   &119.88 \\
    \hline
    \ce{Sr2F}   &2.13&2.13     &114.4    \\
    \ce{Ca2F}   &2.26&2.26     &117.2   \\
    \hline
  \end{tabular}
\end{table}
\begin{table}
  \caption{Stationary points on the potential energy surface of the electronic ground state optimized using the CCSD(T) method for SrFLi, SrFRb (${}^1A'$), and \ce{Sr2F} (${}^2A'$) trimers. $GM$, $LM$, $SP$ superscripts correspond to global- and local minima, and  saddle  point, respectively.  The bond lengths are given in \AA.}
  \label{tbl:lowspin_points}
  \begin{tabular}{llllllllllll}
    \hline
     System & R${}^{GM}_{SrF}$ & R${}^{GM}_{FX}$ & $\theta_{SrFX}^{GM}$  && R${}^{SP}_{SrF}$ & R${}^{SP}_{FX}$ & $\theta_{SrFX}^{SP}$ && R${}^{LM}_{SrF}$ & R${}^{LM}_{FX}$ & $\theta_{SrFX}^{LM}$  \\
    \hline
    SrFLi &2.27 &1.70 &109.73  && 2.11 & 3.27 & 69.1 && 2.13 & 5.61 & 0 \\
    SrFRb &2.26 &2.60  &120.45  && 2.12& 4.41 & 65.75 && 2.17 & 6.55 & 0 \\
    \ce{Sr2F} & 2.26 & 2.26 & 116.42  && 4.58 & 2.10 & 66.89 &&2.15 & 2.75 & 105.24  \\
    \hline
  \end{tabular}
\end{table}
\subsection{Potential energy surfaces in Jacobi coordinates for SrF interacting with Li, Rb and Sr}
For the future scattering calculations it is crucial to know how reactants and products interact in the entrance and exit channels.
To this end, we have studied interaction energy surface in Jacobi coordinates for SrF+Li, SrF+Rb, and SrF+Sr systems.
Interatomic distances in the SrF, LiF, and RbF molecules were fixed to experimental values 
 ~\cite{Huber}. They read 2.075, 1.56 and 2.27 \AA{} for  SrF, LiF and RbF molecules, respectively.

Figure ~\ref{fgr:srfli} shows the potential energy surface for SrF+Li and LiF+Sr interactions in the singlet electronic ground state. Both potentials are very deep and strongly anisotropic, with the well depths on the order of thousands of cm$^{-1}$. For the fixed  SrF bond length we found two minima separated by the energy barrier which is  below zero,
so that both minima are accessible for collisions with the low collision energy. On the Sr+LiF potential energy surface we found one, very deep  minimum. Both global minima on the SrF+Li and  LiF+Sr surfaces converge to the global minimum  reported in previous section after allowing SrF/LiF bonds to relax. Hence, no barrier exists between SrF+Li and LiF+Sr chemical arrangements. A very similar behavior is can be seen for the SrF+Rb  and RbF+Sr  systems (cf. Figure ~\ref{fgr:srfrb}): both potential energy surfaces are very deep and extremely anisotropic, there are two minima on the SrF+Rb  and one minimum on the RbF+Sr surface. Again, both minima on the surface in the Jacobi coordinates originate from  the global minimum for the SrF+Rb system which has a similar geometry. This leads us to the conclusion that no activation barrier exist for the RbF+Sr$\to$ SrF+Rb reaction.  Note also, that spin-restricted single-reference calculations are valid in quite limited range of geometries  (up to about  6 \AA) and the RHF and CC calculations fail to converge for larger distances between atom and the COM of a molecule. Similarly, the SrF+Sr potential energy surface is very strongly anisotropic although the well depth of such surface is  significantly shallower compared to the interactions with alkali-metal atoms.

We also used  supermolecular  calculations to evaluate the asymptotic part of the interaction potential. The long-range interactions are particularly important in studies of cold collisions.  To describe correctly the long-range forces we calculated the leading van der Waals coefficients from the potential obtained using the spin-unrestricted UCCSD(T) method \cite{Bartlett:1993,Werner:1993}. Since singlet and triplet electronic states of SrF+alkali-metal-atom species are asymptotically degenerate, and the former system is single-reference such the procedure is valid.
From the tail of the isotropic and anisotropic parts of the interaction potential we have evaluated  the  van der Waals $C_{6,0}$ and $C_{6,2}$ coefficients which are collected in the Tab. ~\ref{tbl:c6}.
The isotropic van der Waals  coefficient $C_{6,0}$  in the  SrF+Li system is  significantly smaller than in SrF+Rb and SrF+Sr systems, which can be rationalized by comparing  polarizabilities of the Li atom (161 a.u.) to Rb (319 a.u.) and Sr (195 a.u.) atoms. Interestingly enough, the $C_{6,0}$ coefficients obtained here are much smaller than those for alkali-metal-atoms trimers \cite{Zuchowski:2014}. 
The anisotropic coefficients $C_{6,2}$ for all systems are order of magnitude smaller than $C_{6,0}$, they read 135, 288, 152 a.u. for SrF+Li, SrF+Rb, SrF+Sr, respectively.

\begin{table}
  \caption{The  isotropic and anisotropic van der Waals coefficients (in atomic units) for the interactions of the SrF molecule with Li, Rb, and Sr atoms obtained by fitting to the long-range part of the UCCSD(T) potentials projected on the Legendre polynomials $P_0$ and $P_2$ .}
  \label{tbl:c6}
    \begin{tabular}{lllll}

    \hline
    System  & $C_{6,0}$ &$C_{6,2}$ \\
    \hline
    SrF+Li & 1880  &135\\
    SrF+Rb & 3495 &288\\
    SrF+Sr & 2631 &152\\
    \hline
  \end{tabular}
\end{table}

Strong anisotropy of the potential energy surface might have important consequences for the prospects of sympathetic cooling of the system. All quantum numbers of the SrF molecule can be coupled to potential anisotropy through the molecular rotational quantum number and the end-over-end angular momentum. If such molecules are prepared in the excited states (rovibrational, hyperfine, low-field-seeking Zeeman or Stark  states) their collisions with ultracold Rb, Li or Sr atoms might be dominated by inelastic ones due to a very strong couplings in the system. 
There is a hope that the rates of such unfavorable inelastic collisions might be suppressed in case of the SrF+Li system due to higher centrifugal barrier~\cite{Wallis2011}: for the SrF+Li system the  $p$- and $d$-wave centrifugal barrier  heights are  2 mK and 11mK high, respectively,  compared to 70 $\mu$K  and 400 $\mu$K  for the SrF+Rb system. Nonetheless, further scattering calculations (involving also high-spin potentials in case of alkali-metal+SrF) are essential to elucidate this issue.  One has to mention, that such calculations are by no means easy, since the potential anisotropy makes the close-coupling calculations prohibitively expensive, very large rotational basis sets are needed to converge calculations for such strongly anisotropic system. A solution to this problem might be application of the total-angular-momentum representation for the calculations in external fields~\cite{Tscherbul:2010,Tscherbul2012}.

\begin{figure}
  \caption{Comparison between (a) SrF+Li and (b) LiF+Sr potential energy surfaces of the $^{1}A'$ electronic state computed in Jacobi coordinates. The energy unit is cm${}^{-1}$.}
  \begin{tabular}{cc}
\includegraphics[scale=0.4]{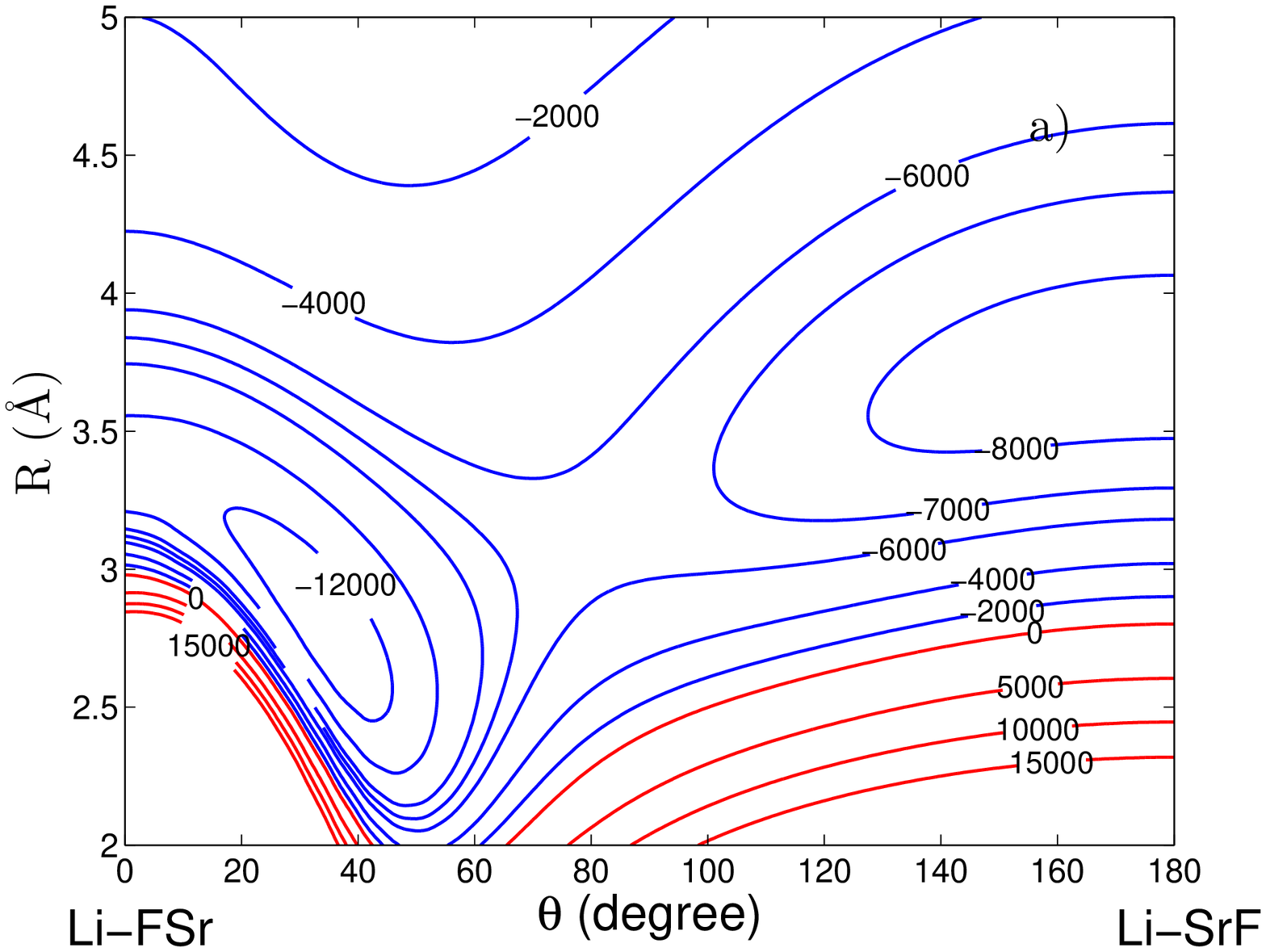} & \includegraphics[scale=0.4]{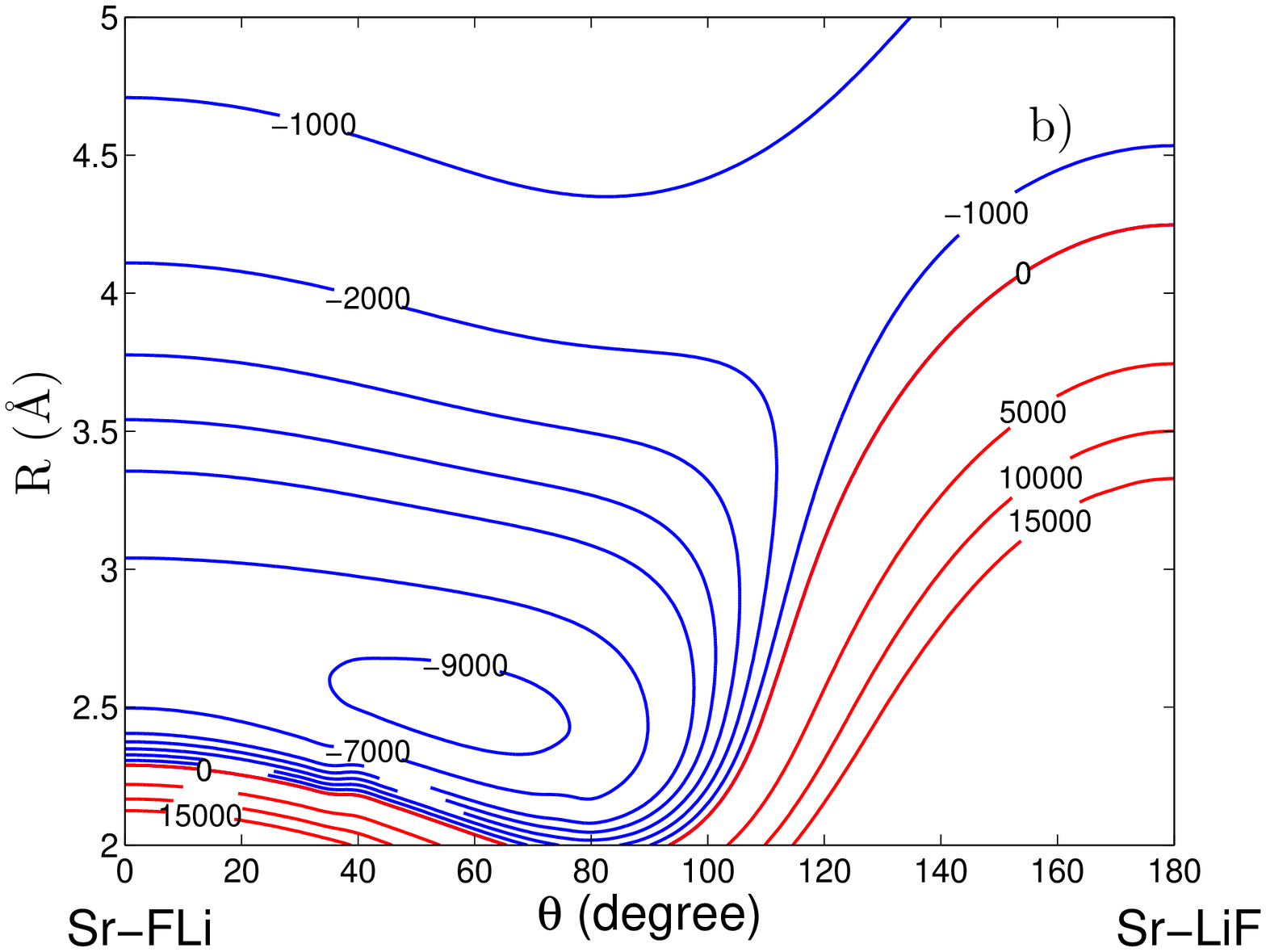}\tabularnewline
\end{tabular}
  \label{fgr:srfli}
\end{figure}

\begin{figure}
  \caption{Comparison between (a) SrF+Rb and (b) RbF+Sr potential energy surfaces of the
$^{1}A'$ electronic state computed in Jacobi coordinates. The energy unit is cm${}^{-1}$.}
  \begin{tabular}{cc}
\includegraphics[scale=0.4]{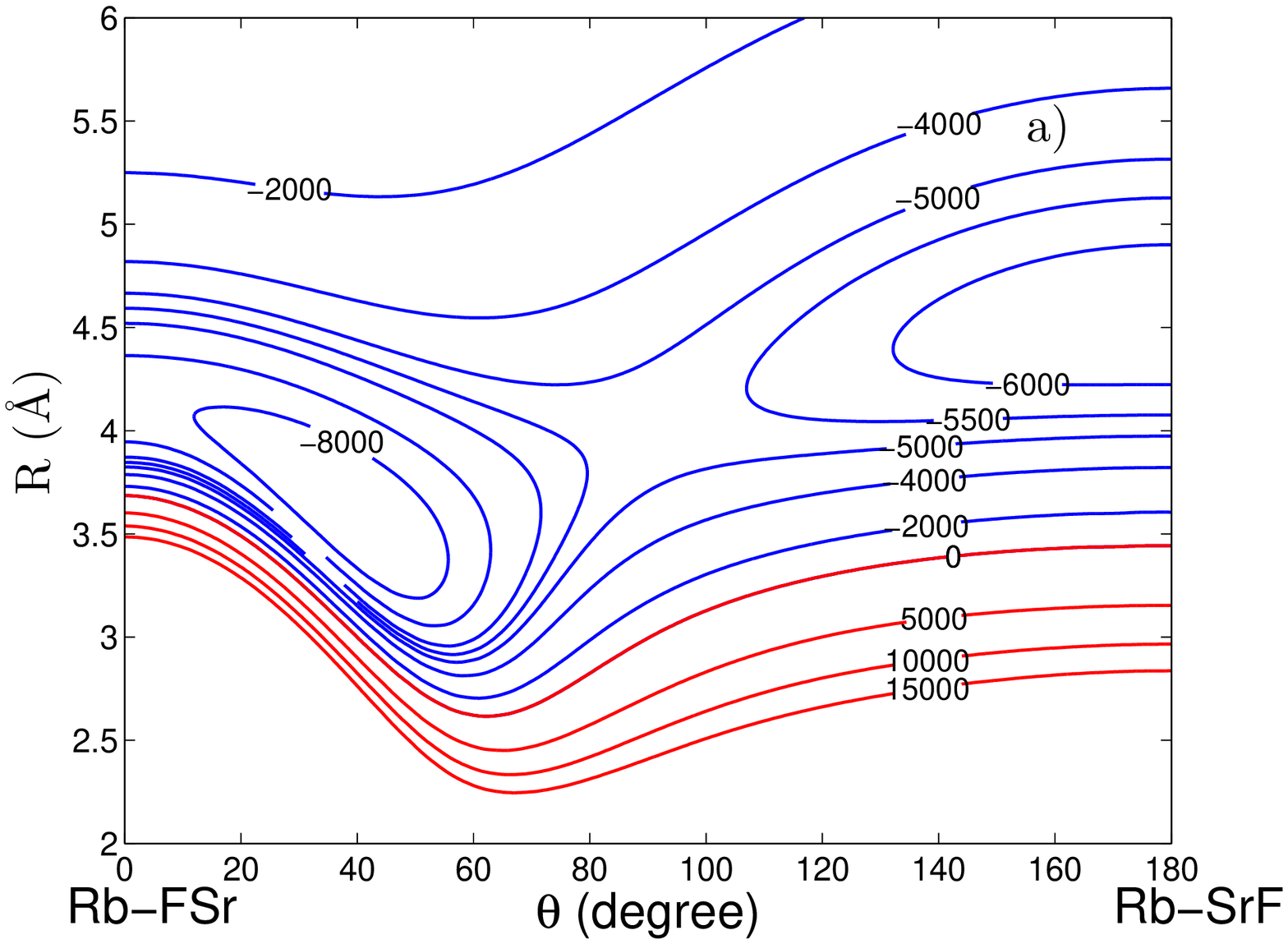} & \includegraphics[scale=0.4]{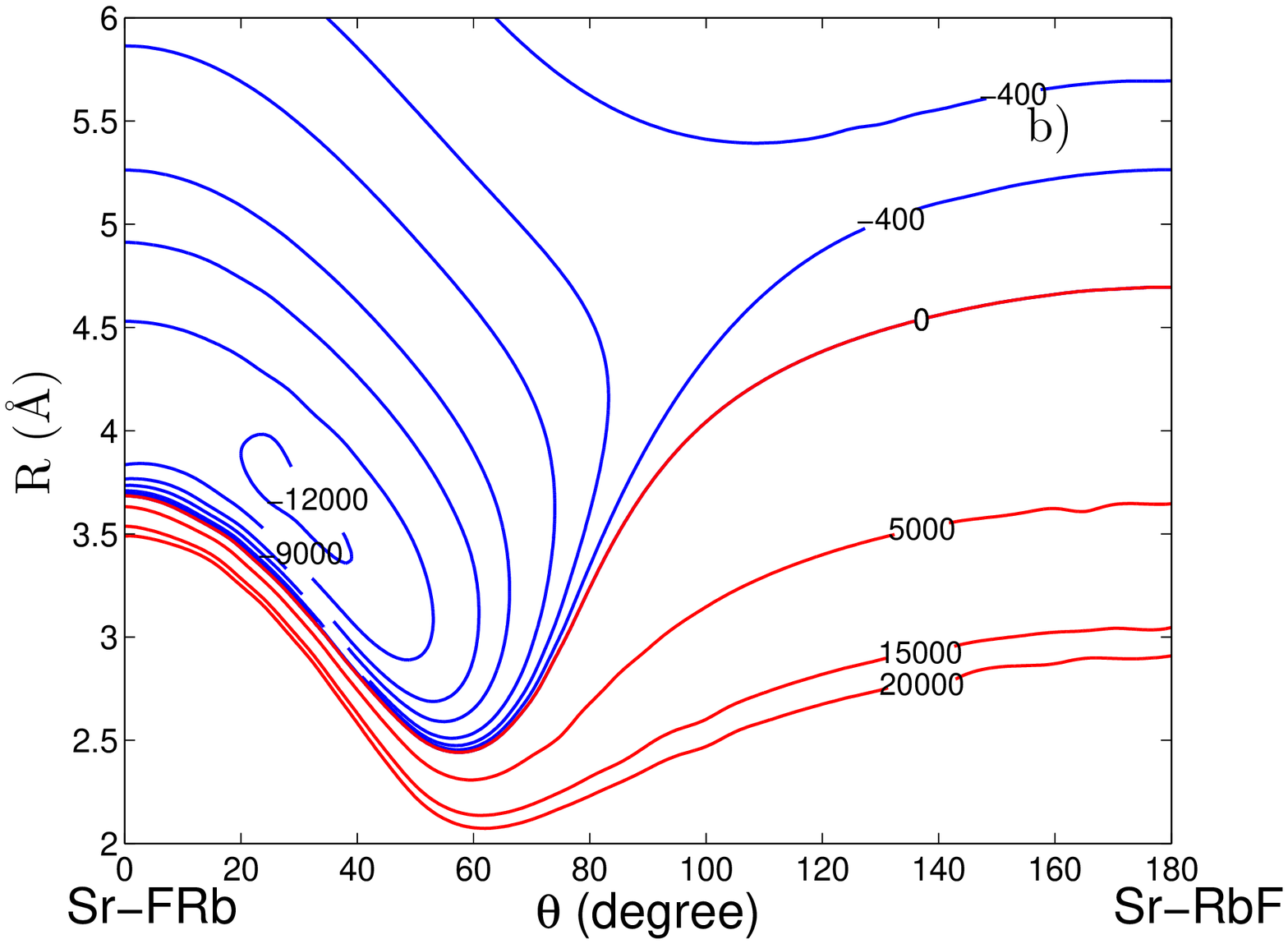}\tabularnewline
\end{tabular}
  \label{fgr:srfrb}
\end{figure}

\begin{figure}
  \caption{SrF+Sr interactions in the $^{2}A'$ electronic ground state in Jacobi coordiantes. The energy unit is cm${}^{-1}$.}
\includegraphics[scale=0.4]{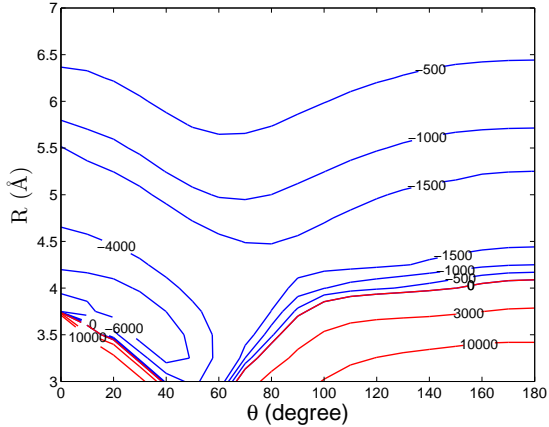}
  \label{fgr:srfsr}
\end{figure}

\subsection{Atom-exchange reaction pathways}

One of the most important questions regarding the collisions of the SrF molecule with alkali-metal   and Sr atoms, is whether the reactions are inhibited by the activation barriers.  
We investigated potential energy surfaces of SrFLi, SrFRb and SrFSr trimers for the exchange reaction.
To this end the internal coordinates were employed: the Sr-F-X (where X is Li, Rb, Sr) angles were fixed, while the Sr-F, and F-X distances were varied.
Fixed angles correspond to global minima of SrFX systems reported in Table~\ref{tbl:lowspin_points}.
Maps of the potential energy surfaces for atom-fluorine separations are shown in Figure \ref{fgr:r1r2}. The interatomic distances were varied up until 4 \AA{} which is about twice the bond length of the SrF molecule.

All potential energy surfaces show a smooth behavior with one minimum and no barriers for the geometries concerned. We experienced no problems using  the RCCSD(T) method and the T1-diagnostic was only slightly larger than recommended value of 0.02.  It is interesting to see, that for the Rb+SrF system the geometries accessible for  the low collision energies  are always very close to the equilibrium distance of the SrF molecule. For the SrF+Sr reaction we found no barrier for the atom exchange. The direction of such reaction might be controlled by proper choice of isotopologues. For example, the $^{88}$SrF+$^{86}$Sr reaction would be energetically forbidden, while the reaction $^{86}$SrF+$^{88}$Sr  could occur without barrier and release small amount of energy that could be on the order of zero point energy change. 
Thus, the energy difference for the ${}^{86}$SrF+${}^{88}$Sr$\to{}^{88}$SrF+${}^{86}$Sr reaction equals 0.518 cm${}^{-1}$.
Similar idea of the isotope exchange in reactions of ultracold heteronuclear dimers was recently proposed by Tomza \cite{Tomza:PRL}.

\begin{figure}
  \caption{Potential energy surfaces for atom exchange in (a) SrF+Li , (b) SrF+Rb systems, both in $^{1}A'$ electronic states, and (c) SrF+Sr system in $^{2}A'$ state. The Sr-F-X angle is fixed and corresponding to the global minimum (see Table \ref{tbl:lowspin_points}).
  The energy unit is cm$^{-1}$. The zeros on the potential energy surfaces correspond to SrF+Li, SrF+Rb and SrF+Sr dissociation channels.} 

\includegraphics[scale=0.5]{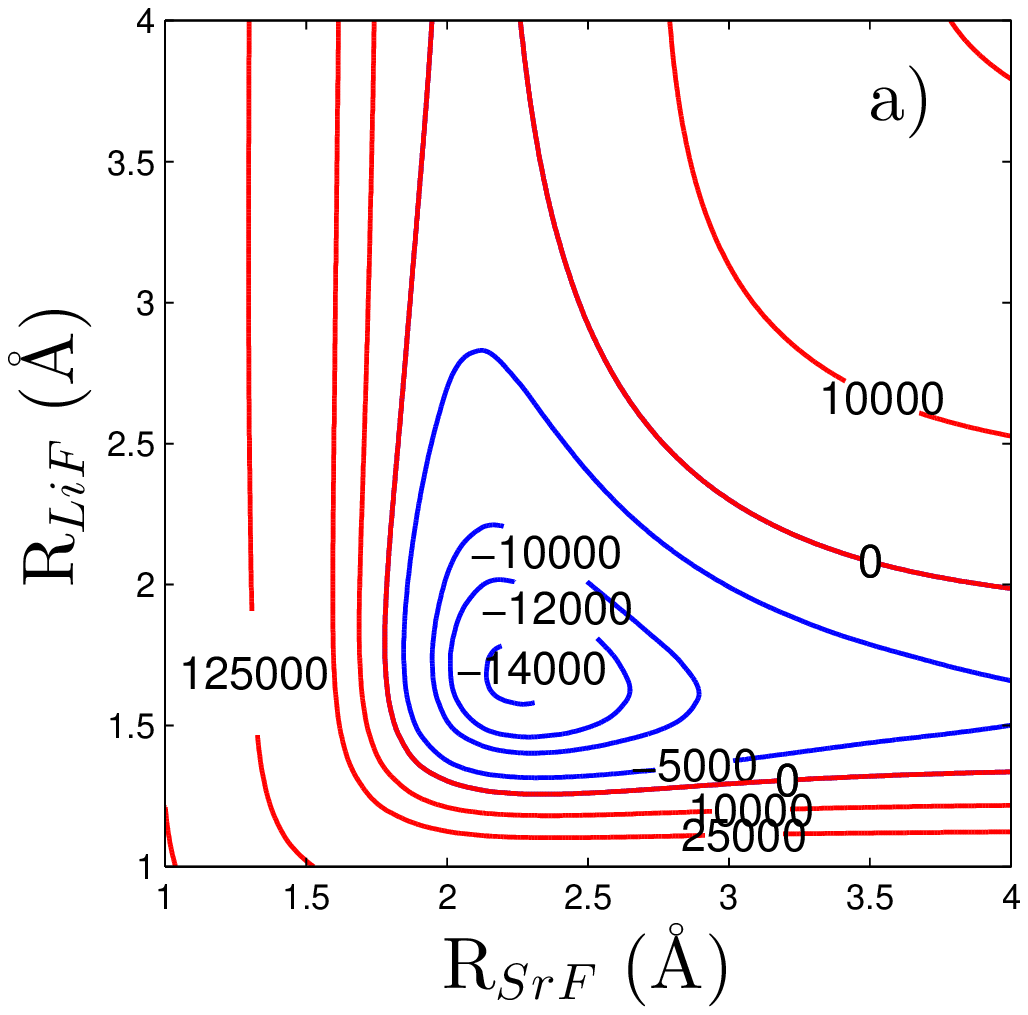} 
\includegraphics[scale=0.5]{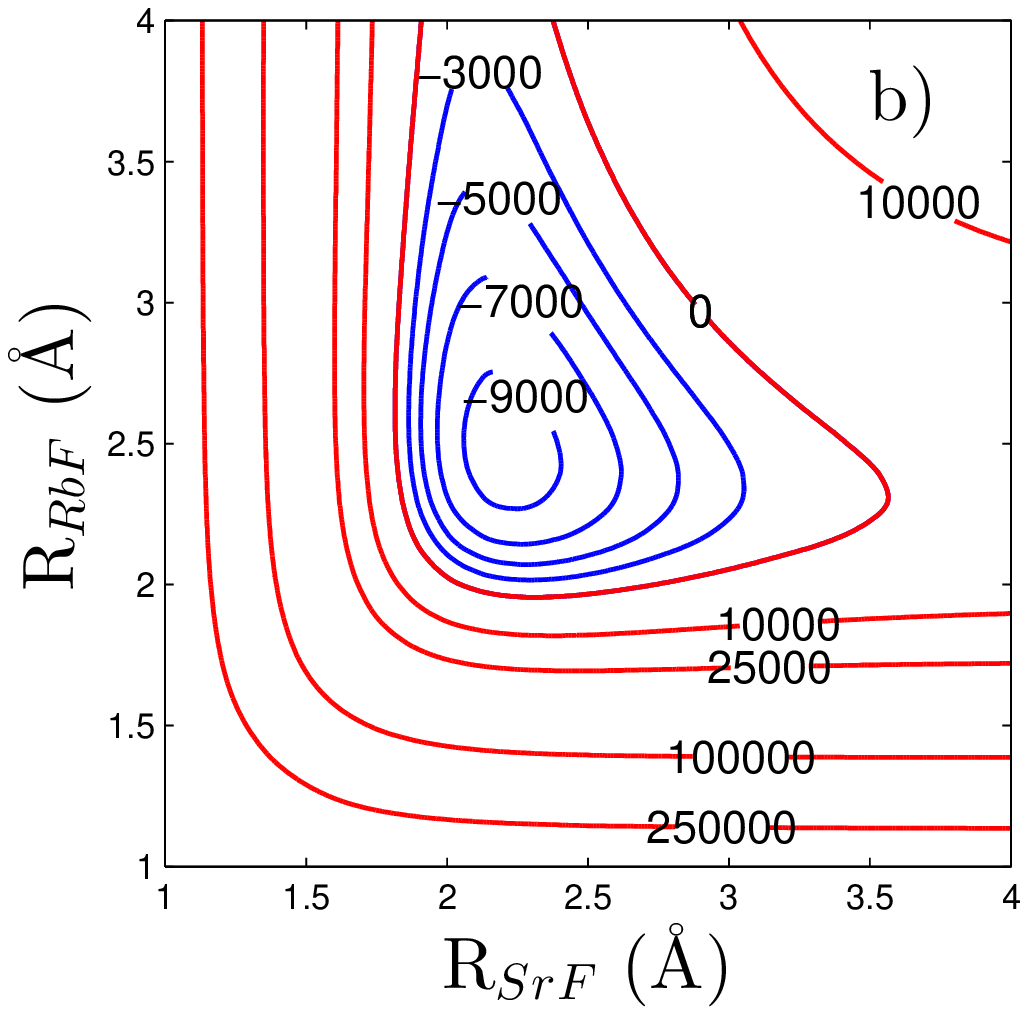} 
\includegraphics[scale=0.5]{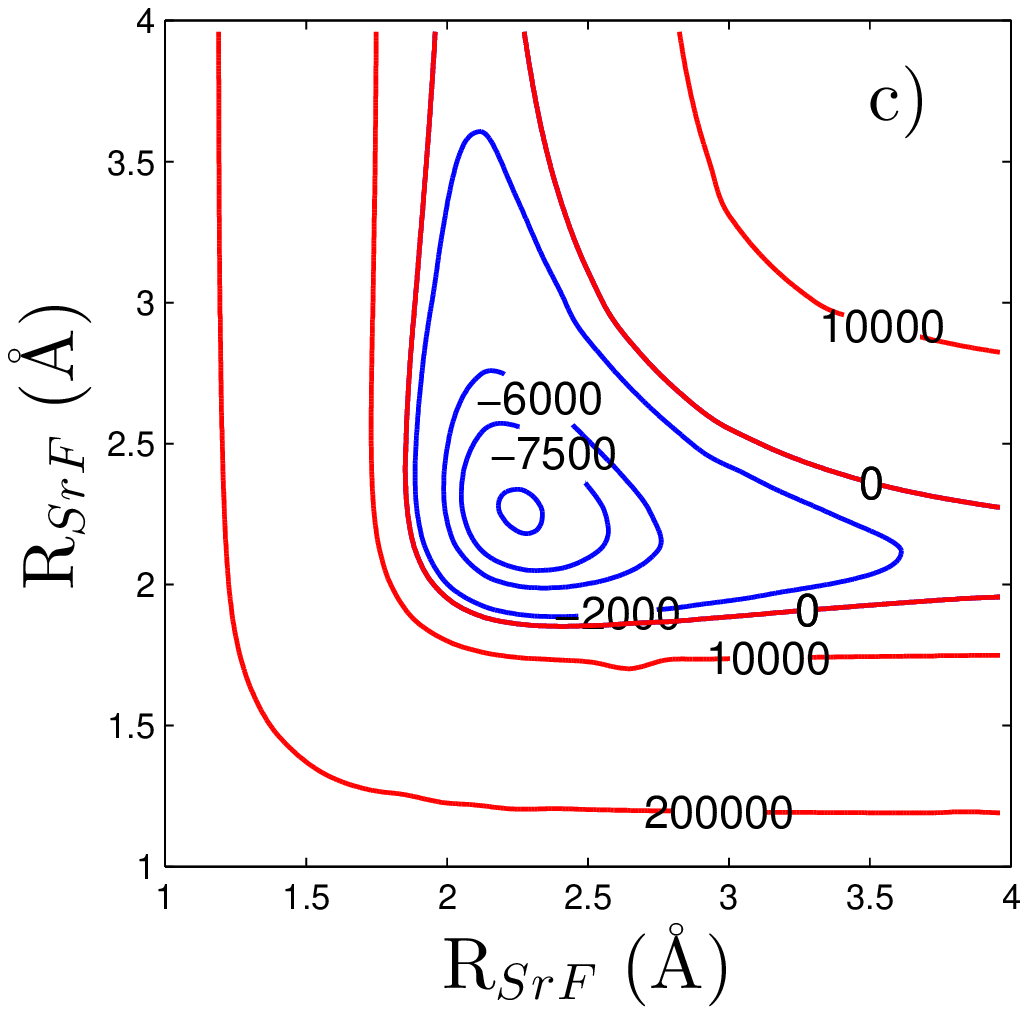} 
  \label{fgr:r1r2}
\end{figure}

\section{Conclusions}
In this work we examined the reactive and non-reactive potential energy surfaces of the SrF+Rb, SrF+Li, and SrF+Sr systems in the lowest spin states. 
We found that collisions with the low kinetic energy can be reactive only for the SrF+Li system. Large excess of energy released in such reaction is enough to populate up to first five vibrational energy levels of the LiF molecule. 
Other alkali metal atoms will not detach the F atom from the SrF molecule and in such case they can be used for the sympathetic cooling of the SrF molecules in their absolute ground state (e.g., in optical or microwave traps \cite{DeMille2004,Tokunaga2011}). 
We also found that the SrF+Li as well as SrF+Sr reactions are barrierless.
For the short atom-molecule distances the potential describing such interactions are very deep.
For the SrF+Rb system the reaction is energetically forbidden, however, the reverse reaction RbF+Sr can go without activation barrier. 
We also calculated the potential energy surfaces for SrF+Li,  SrF+Rb, and SrF+Sr systems in the Jacobi coordinates. These potentials are extremely anisotropic and quite likely, the SrF molecules trapped in excited (rovibrational and hyperfine, Stark or Zeeman) states will undergo collisional quenching.   
To predict the usefulness of the sympathetic cooling of the SrF molecule by alkali metal atoms and alkaline-earth further studies are essential, involving scattering calculations, also in the spin-stretched state, which can be prepared in the magnetic trap. Work in this direction will be continued in future.

\begin{acknowledgement}
The authors acknowledge support from grant no. DEC-2012/07/B/ST4/01347 and the Homing Plus programme (Project No. 2011-3/14) of the Foundation for Polish Science, which is co-financed by the European Regional Development Fund of the European Union.  We are also grateful for the computer time provided by the Wroclaw Centre of Networking and Supercomputing (Project No. 218).
We thank Pawel Tecmer for his helpful comments and Timur Tscherbul for interesting us in this problem. 
\end{acknowledgement}


\bibliographystyle{achemso}

\providecommand{\latin}[1]{#1}
\providecommand*\mcitethebibliography{\thebibliography}
\csname @ifundefined\endcsname{endmcitethebibliography}
  {\let\endmcitethebibliography\endthebibliography}{}

\end{document}